# Acoustic lens associated with a radial oscillating bubble

Ion Simaciu[1,a], Gheorghe Dumitrescu[2], Zoltan Borsos[1,b],

1 Petroleum-Gas University of Ploiești, Ploiești 100680, Romania

2 High School Toma N. Socolescu, Ploiești, Romania

E-mail: [a] isimaciu@yahoo.com; [b] borzolh@upg-ploiesti.ro

**Abstract**

In this paper we show that a radial oscillating bubble in a liquid behaves like a convergent acoustic lens.

1. **Introduction**

In a previous paper [1] we have shown that a packet of spherical stationary acoustic wave acts as a convergent acoustic lens.

In this paper, we follow to investigate some consequences of our approach of the acoustic world. Therefore, we found that also a radial oscillating bubble behaves like an acoustic lens. This is the first issue of this paper. Then, in the second section we infer the expression of the acoustic pressure around a bubble that oscillates radially under the action of a plane wave. Some interesting properties of the refractive index around an oscillating bubble can be found exploring its shape obtained in paper [1]. Our derivation in the third section leads to the dependence of the refractive index on the absolute value of the position vector.

When the focal length of the lens is equal to the absolute value of the position vector then the bubble behaves like a dumb hole. But according to the fourth section of this paper a bubble in liquid cannot become a dumb hole. This conclusion could be transferred to the electromagnetic world in order to describe the behaviour of the electron. The fifth section is devoted to the discussions and conclusions.

2. **Acoustic pressure around an oscillating bubble**

To date the radial oscillation of a bubble is already investigated theoretically [2-7] and experimentally [8, 9]. This kind of oscillation must be assumed to a wave which exerts an oscillating pressure.

Given a bubble of radius $R_0$ at equilibrium, immersed in a fluid with hydrostatic pressure $p_0$. We will assume that this bubble is forced to oscillate under the action of a pressure wave with constant amplitude $A$ and angular frequency $\omega$

$$p_{ext}(t) = p_0 + A\cos\omega t. \tag{1}$$

Under the action of this wave, the bubble performs radial oscillations with small amplitude

$$R(t) = R_0\left[1 + a\cos(\omega t + \varphi)\right]. \tag{2}$$

According to Rayleigh-Plesset equation [2-4], the dimensionless amplitude and the forced oscillation phase are

$$a = \frac{A}{\rho R_0^2\left[\left(\omega^2 - \omega_0^2\right)^2 + 4\beta^2\omega^2\right]^{1/2}}, \quad \varphi = \arctan\frac{2\beta\omega}{\left(\omega^2 - \omega_0^2\right)}. \tag{3}$$

Here (3), $\omega_0$ is the natural angular frequency of the bubble and $\beta$ the radial damping constant.

The pressure around the bubble, due to bubble volume oscillations, is

$$p'(r,t) = \frac{\rho\ddot{V}}{4\pi r} - \frac{\rho\dot{R}^2}{2}\left(\frac{R}{r}\right)^4 \cong \frac{\rho\ddot{V}}{4\pi r} = \frac{\rho R}{r}\left(2\dot{R}^2 + R\ddot{R}\right) \cong \frac{\rho R^2\ddot{R}}{r}. \tag{4}$$

Following (2), (3) and (4), yields

$$p'(r,t) \cong -\frac{R_0}{r}\frac{\omega^2 A\cos(\omega t + \varphi)}{\left[\left(\omega^2 - \omega_0^2\right)^2 + 4\beta^2\omega^2\right]^{1/2}}, \quad r \geq R_0. \tag{5}$$

At resonance, for angular frequency close to natural angular frequency $\omega \cong \omega_0$, the relationship (5) becomes

$$p'_{res}(r,t) \cong -\frac{R_0}{r}\frac{\omega_0 A\cos(\omega_0 t + \varphi_0)}{2\beta_0}, \quad r \geq R_0, \tag{6}$$

with

$$\beta_0 = \beta_{0ac} + \beta_{0\mu} \cong \beta_{0ac} = \frac{\omega_0^2 R_0}{2u} \tag{7}$$

and, if $\beta_{0ac} \gg \beta_{0\mu}$, then

$$\varphi_0 = \arctan\left[\lim_{\omega\to\omega_0}\frac{2\beta\omega}{\left(\omega^2 - \omega_0^2\right)}\right] = \arctan\infty = \frac{\pi}{2}. \tag{8}$$

When (7) and (8) will be subtracted in the expression of pressure (6), it results

$$p'_{res}(r,t) \cong -\frac{A}{r}\frac{u\sin(\omega_0 t)}{\omega_0}, \quad r \geq R_0. \tag{9}$$

In relations (7) and (9), $u$ is the acoustic wave velocity in the undisturbed liquid.

According to the relations (5) and (6), an oscillating pressure field is established around the bubble. It depends on the position, relative to the center of the bubble.

The temporal average of the pressure is zero ($\langle p'(r,t)\rangle_t = 0$. However, according to the dynamics of the waves in a compressible fluid [10, Ch.8-§64], the variation of the average density around the bubble is not zero. This nonzero variation of the average density implies a change of the velocity with which waves travel the medium. This change leads therefore to the dependence of the acoustic refractive index on the position.

## 3. The refractive index for a radially oscillating bubble

In following we will use the refractive index [1] for inhomogeneous medium

$$\langle n_{ab}(p) \rangle_t \cong 1 + \frac{\zeta(3\zeta+1)}{2(2\zeta+1)^2} \frac{\langle (\delta p)^2 \rangle_t}{p_0^2}. \tag{10}$$

Subtracting (5) and (9) in relation (10), and setting $p'(r,t) = \delta p(r,t)$, yields

$$\langle n_{ba}(p) \rangle_t \cong 1 + \frac{\zeta(3\zeta+1)}{4(2\zeta+1)^2} \frac{R_0^2}{r^2} \frac{A^2}{p_0^2} \frac{\omega^4}{\left[(\omega^2-\omega_0^2)^2 + 4\beta^2\omega^2\right]} > 1, \tag{11}$$

$$\langle n_{ba}(p_{res}) \rangle_t \cong 1 + \frac{\zeta(3\zeta+1)}{8(2\zeta+1)^2} \frac{u^2}{\omega_0^2 r^2} \frac{A^2}{p_0^2} > 1. \tag{12}$$

The dependence on the absolute value of the position vector of the acoustic refractive index reveals the acoustic inhomogeneity, with spherical symmetry, of the liquid around the oscillating bubble. That is the liquid surrounding the bubble behaves like a spherical lens.

Given an acoustic wave which is plane, and which is deflected when it passes at the minimum distance from the center of the bubble. The focal length of the spherical lens is [11-Ch. 27.3]

$$\frac{1}{f} = \frac{2}{r}\left(\langle n_{ba}\rangle_t - 1\right) = \frac{\zeta(3\zeta+1)}{2(2\zeta+1)^2} \frac{R_0^2}{r^3} \frac{A^2}{p_0^2} \frac{\omega^4}{\left[(\omega^2-\omega_0^2)^2 + 4\beta^2\omega^2\right]} > 0 \tag{13a}$$

or

$$f(r) = \frac{2(2\zeta+1)^2}{\zeta(3\zeta+1)} \frac{r^3}{R_0^2} \frac{p_0^2}{A^2} \frac{\left[(\omega^2-\omega_0^2)^2 + 4\beta^2\omega^2\right]}{\omega^4} > 0, r \geq R_0, \tag{13b}$$

that is, the spherical lens is convergent.

When resonance take place, the focal length becomes

$$f_{res}(r) = \frac{4(2\zeta+1)^2}{\zeta(3\zeta+1)} \frac{r^3\omega_0^2}{u^2} \frac{p_0^2}{A^2} > 0, r \geq R_0 \tag{14}$$

according to (12).

## 4. Can a bubble become a dumb hole?

The convergent acoustic lens, associated with a bubble, behaves like a dumb hole if

$$f(r) = r. \tag{15}$$

Then, according to (13b), the corresponding dumb hole radius becomes

$$r_{ba} = R_0 \frac{\sqrt{\zeta(3\zeta+1)}}{\sqrt{2}(2\zeta+1)} \frac{A}{p_0} \frac{\omega^2}{\sqrt{(\omega^2-\omega_0^2)^2 + 4\beta^2\omega^2}} \ll R_0. \tag{16}$$

At resonance, the radius of the dumb hole is

$$r_{ba,res} = \frac{\sqrt{\zeta(3\zeta+1)}}{2(2\zeta+1)} \frac{uA}{\omega_0 p_0}. \quad (17)$$

At equilibrium, the pressure of a liquid $p_0$, is [1, 10-Ch.8-§63 ] $(2\zeta+1)p_0 = \rho u^2$ and the angular frequency is [3] $\omega_0 = \{3\gamma[p_0/(\rho R_0^2) + 2\sigma/(\rho R_0^3)] - 2\sigma/(\rho R_0^3)\}^{1/2} = [p_{eff}/(\rho R_0^2)]^{1/2}$. Therefore [17] maybe expressed as

$$r_{ba,res} = R_0 \sqrt{\frac{\zeta(3\zeta+1)}{4(2\zeta+1)}} \frac{A}{\sqrt{p_{eff}\, p_0}}. \quad (18)$$

Since $A \ll p_0$, then $r_{ba} \ll R_0$, i.e. the bubble deflects acoustic waves without capturing them. This happens when waves pass to the minimum distance $r \geq R_0$. That is a bubble cannot become a dumb hole.

One can observe a similar behaviour of the electron when the radius of the cross section of the electromagnetic radiation, $r_e \cong e^2/(m_e c^2)$, is much larger than the gravitational radius, $r_{ge} = 2Gm_e/c^2$.

## 5. Discussions

As we already mentioned above, we have followed paper [1] in this paper in order to find out other consequences which occur when a wave packet travels in a fluid. Therefore, we find out that around an oscillating bubble the fluid becomes inhomogeneous.

Bubble behaves like a convergent spherical lens that deflects the acoustic waves, regardless of the phase of the radial oscillation.

Hence one can construe that the oscillating bubble has both electroacoustic charge [12] and acoustic gravitational charge, that is, it attracts the acoustic waves that propagate at the side of it.

In a further paper we will attempt to prove that two bubbles attract each other by of force which is independent of the oscillation phases.

According to the fourth section the acoustic radius is approximately equal to the radius of the bubble, $r_{ba} \simeq R_0$, if the wave amplitude of the wave, which supply the oscillation, is close to the value of the fluid pressure $A \simeq p_0$. This constraint may lead to a further investigation of the bubble oscillation with large amplitude.

Another case which may be interesting to study is how a spherical cluster of bubbles oscillates when $R_c > R_0$. We assume that bubbles can associate in a cluster, when they interact. The mentioned study must be addressed to the spherical pressure field around the bubble and to the corresponding refractive index. We presume that from this investigation may arise a state of the cluster when the cluster radius is equal to acoustic radius $r_{ca} = R_c$.